\documentclass[reprint,aps,prd,groupedaddress,nofootinbib]{revtex4-1}

\usepackage{dcolumn}
\usepackage{amsmath}
\usepackage{amssymb}
\usepackage{bm}
\usepackage{graphicx}

\usepackage{multirow}

\usepackage{color}
\usepackage{ulem}

\usepackage{hyperref}

\begin{document}

\title{Optical and mechanical properties of ion-beam-sputtered MgF$_2$ thin films \\for gravitational-wave interferometers}

\author{M. Granata}
	\email[]{m.granata@lma.in2p3.fr}
\author{D. Forest}
\affiliation{Laboratoire des Mat\'{e}riaux Avanc\'{e}s - IP2I, CNRS, Universit\'{e} de Lyon, Universit\'{e} Claude Bernard Lyon 1, F-69622 Villeurbanne, France}

\author{A. Amato}
\author{G. Cagnoli}
\affiliation{Universit\'{e} de Lyon, Universit\'{e} Claude Bernard Lyon 1, CNRS, Institut Lumi\`{e}re Mati\`{e}re, F-69622 Villeurbanne, France}

\author{M. Bischi}
	\email[]{matteobischi92@gmail.com}
\author{F. Piergiovanni}
\author{F. Martelli}
\author{M. Montani}
\author{G. M. Guidi}
\affiliation{Universit\`{a} degli Studi di Urbino Carlo Bo, Dipartimento di Scienze Pure e Applicate, I-61029 Urbino, Italy}
\affiliation{INFN, Sezione di Firenze, I-50019 Sesto Fiorentino, Firenze, Italy}

\author{M. Bazzan}
\author{G. Favaro}
\author{G. Maggioni}
\affiliation{Dipartimento di Fisica e Astronomia, Universit\`{a} degli Studi di Padova, I-35131, Padova, Italy}

\author{F. Schiettekatte}
\author{M. Chicoine}
\affiliation{Universit\'{e} de Montr\'{e}al, Montr\'{e}al, Qu\'{e}bec, Canada}

\author{M. Menotta}
\affiliation{Universit\`{a} degli Studi di Urbino Carlo Bo, Dipartimento di Scienze Biomolecolari, I-61029 Urbino, Italy}

\author{A. Di Michele}
\affiliation{Universit\`{a} degli Studi di Perugia, Dipartimento di Fisica e Geologia, Via Pascoli, 06123 Perugia, Italy}

\author{M. Canepa}
\affiliation{OPTMATLAB, Dipartimento di Fisica, Universit\`{a} di Genova, Via Dodecaneso 33, 16146 Genova, Italy}
\affiliation{INFN, Sezione di Genova, Via Dodecaneso 33, 16146 Genova, Italy}

\date{\today}

\begin{abstract}
Brownian thermal noise associated with highly reflective coatings is a fundamental limit for several precision experiments, including gravitational-wave detectors. Research is currently ongoing to find coatings with low thermal noise that also fulfill strict optical requirements such as low absorption and scatter. We report on the optical and mechanical properties of ion-beam-sputtered magnesium fluoride thin films, and we discuss the application of such coatings in current and future gravitational-wave detectors.
\end{abstract}

\pacs{PACS}

\keywords{Keywords}

\maketitle

\section{Introduction}
Brownian thermal noise in highly reflective coatings \cite{Saulson90,Levin98} is a fundamental limitation for precision experiments such as interferometric gravitational-wave detectors \cite{Adhikari14}, optomechanical resonators \cite{Aspelmeyer14}, and frequency standards \cite{Matei17}. As measured with a laser beam, its power spectral density $S$ can be written as \cite{Harry02}
\begin{equation}\label{eqn.S}
S \propto \frac{k_B T}{2\pi f} \frac{d}{w^2} \varphi_c\ ,
\end{equation}
where $k_B$ is the Boltzmann constant, $f$ is the frequency, $T$ is the temperature, $d$ is the coating thickness, $w$ is the laser beam radius where intensity drops by 1/e$^2$, and $\varphi_c$ is the coating loss angle. The latter quantifies the dissipation of mechanical energy in the coating and is in turn a function of frequency and temperature, $\varphi_c(f, T)$. Thermally induced fluctuations of coated surfaces can thus be reduced by increasing the beam radius, by decreasing the temperature, or by choosing coating materials which minimize the $d\varphi_c$ term in Eq.(\ref{eqn.S}).

High-reflection coatings are usually Bragg reflectors of alternating layers of high and low refractive indices $n_{\textrm{\tiny{H}}}$ and $n_{\textrm{\tiny{L}}}$, respectively. For the same coating transmissivity, the thickness of the layers and the number of layer pairs are a monotonically decreasing function of the refractive index contrast $C = n_{\textrm{\tiny{H}}}/n_{\textrm{\tiny{L}}}$. Thus, the higher the contrast $C$, the lower the high-reflection coating thickness $d$ and hence the coating thermal noise.

The high-reflection coatings of the Advanced LIGO \cite{aLIGO}, Advanced Virgo \cite{AdVirgo} and KAGRA \cite{KAGRA} gravitational-wave detectors are thickness-optimized stacks \cite{Villar10} of ion-beam-sputtered (IBS) layers of tantalum pentoxide (Ta$_2$O$_5$, also known as {\it tantala}, high index) and silicon dioxide (SiO$_2$, {\it silica}, low index), produced by the Laboratoire des Mat\'{e}riaux Avanc\'{e}s (LMA) \cite{Pinard17,Degallaix19}. Following a procedure developed by the LMA to reduce their optical absorption and loss angle \cite{Harry07}, the high-index layers of Advanced LIGO and Advanced Virgo also contain a significant amount of titanium dioxide (TiO$_2$, {\it titania}) \cite{Granata20}. Despite the superb optical and mechanical properties of their current coatings \cite{Degallaix19,Amato19,Granata20}, coating thermal noise remains a severe limitation for further sensitivity improvement in current gravitational-wave detectors. Thus, in the last two decades, a considerable research effort has been committed to find alternative coating materials featuring extremely low mechanical and optical losses (absorption, scatter) at the same time \cite{Granata20review,Vajente21}.

The motivation to find alternative coating materials is even stronger for cryogenic gravitational-wave detectors, either present or future, such as KAGRA, Einstein Telescope \cite{ET1,ET2}, and Cosmic Explorer \cite{Abbott17}. Although the low-temperature behavior of the loss angles of tantala and titania-tantala coatings is still matter of debate to date \cite{Martin09,Martin10,Granata13,Hirose20}, the coating loss angle of silica has been conclusively shown to considerably increase below 30 K \cite{Martin14,Granata15}.

Because of their low refractive index \cite{Allen90,Zukic90,Kolbe92,Kolbe93,Bosch00,Quesnel00,Dumas02,Ristau02,Gunster05,Yoshida06,Yu07,Putkonen11,deMarcos17},   fluoride coatings represent an interesting option to decrease the thickness and hence the thermal noise of the high-reflection coatings of gravitational-wave detectors. Furthermore,  because of their potentially low mechanical loss at low temperature \cite{Schwarz11}, fluorides could be a valid option especially for use at cryogenic temperatures. So far, however, fluoride coatings have never been considered for implementation in gravitational-wave detectors, so that a specifically oriented characterization of their properties is needed.

As a first step towards the development of coatings with low losses, in this paper we report on the optical and mechanical properties of IBS magnesium fluoride (MgF$_2$) thin films measured at ambient temperature, and we discuss their use in gravitational-wave detectors in place of their current low-index silica layers. As post-deposition annealing is a standard procedure to decrease coating loss angle and optical absorption, we took special care to characterize its effect on the coating properties.

\section{Methods}
\subsection{Samples}
$\sim$200-nm thick layers of IBS MgF$_2$ have been deposited on different substrates: (i) silicon wafers ($\varnothing$ 75 mm, $t = 0.5$ mm) for optical characterization, ion beam analysis and x-ray diffraction measurements, and (ii) fused-silica disks ($\varnothing$ 50 mm, $t = 1$ mm) for mechanical characterization. Prior to deposition, the fused-silica disks were annealed in air at 900 $^\circ$C for 10 hours to release their internal stress due to manufacturing and minimize their intrinsic loss angle $\varphi_0$ \cite{Travasso07}.

\begin{figure}
\centering
	\includegraphics[width=0.45\textwidth]{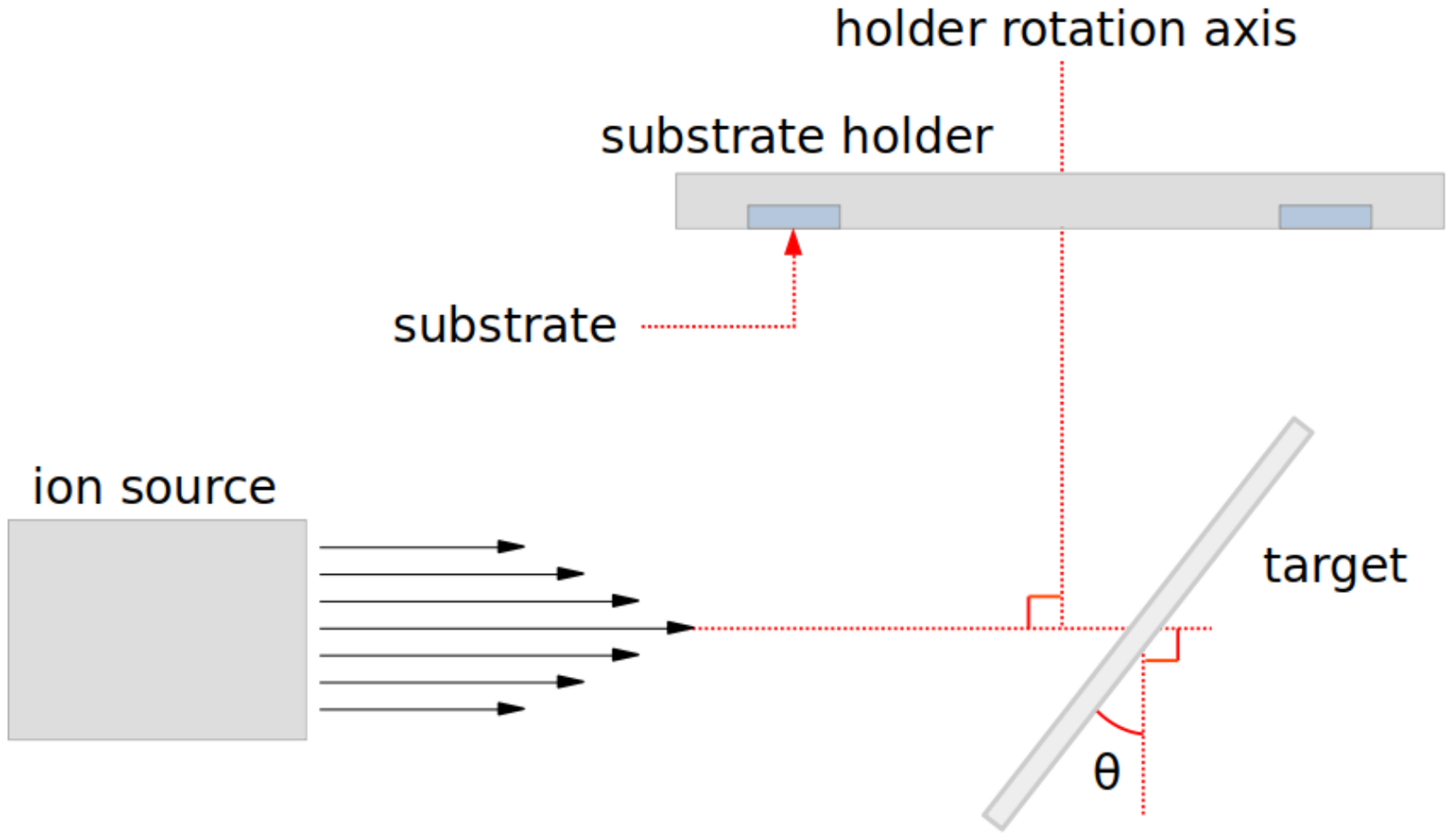}
	\caption{Schematic view of the IBS system used to produce MgF$_2$ thin films, where $\theta=38^{\circ}$.}
	\label{FIG.IBS}
\end{figure}

Coatings were deposited by Laser Zentrum Hannover\footnote{www.lzh.de} via IBS. Prior to deposition, the base pressure inside the coater vacuum chamber was $5 \times 10^{-6}$ mbar. The total pressure during the coating process was $2 \times 10^{-4}$ mbar, with 54 sccm of noble gases (mainly Xe) and gases containing fluorine injected into the chamber. Energy and current of the sputtering ions were 0.9 keV and 0.2 A, respectively, for an average coating deposition rate of 0.1 nm/s. Figure \ref{FIG.IBS} schematically illustrates the layout of the ion beam source, the sputtering target and the rotating substrate holder inside the coater vacuum chamber.

All samples were treated together in a first coating run. Then, in order to cancel out the coating-induced curvature that would affect their mode frequencies, the fused-silica disks underwent a second coating run on their other side, under identical conditions. 

In order to minimize coating mechanical loss $\varphi_c$ and optical absorption $\alpha$, coated samples were thermally treated. The annealing treatments were performed in Ar atmosphere, at overpressure with respect to the environment, to avoid surface oxidation. We tested different soaking temperatures $T_a$ and, with the fused-silica disks only, also different times $\Delta t_a$. More specifically, a disk A underwent a series of treatments of increasing soaking temperature, each one of the same duration, while a disk B underwent a series of treatments of increasing time, each one performed at the same soaking temperature. Parameters used for the annealing runs of the fused-silica disks are summarized in Tables \ref{TABLEannA} and \ref{TABLEannB}.

In between measurements, all samples were stored under primary vacuum ($10^{-2} - 10^{-1}$ mbar) to mitigate oxidation from air exposure.
\begin{table}
\caption{\label{TABLEannA} Soaking temperature $T_a$ and time $\Delta t_a$ of annealing treatments applied to fused-silica disk A. Heating and cooling ramps of 100 $^{\circ}$C/hour were used.}
\begin{ruledtabular}
\begin{tabular}{lccccc}
	& \#1 & \#2 & \#3 & \#4 & \#5\\ \cline{2-6}
	$T_a$ [$^{\circ}$C]& 120 & 200 & 285 & 311 & 373\\
	$\Delta t_a$ [h] & 10 & 10 & 10 & 10 & 10\\
\end{tabular}
\end{ruledtabular}
\end{table}
\begin{table}
\caption{\label{TABLEannB} Soaking temperature $T_a$ and time $\Delta t_a$ of annealing treatments applied to fused-silica disk B. Heating and cooling ramps of 100 $^{\circ}$C/hour were used.}
\begin{ruledtabular}
\begin{tabular}{lccccc}
	& \#1 & \#2 & \#3 & \#4 &\\ \cline{2-5}
	$T_a$ [$^{\circ}$C]& 285 & 285 & 285 & 285\\
	$\Delta t_a$ [h] & 10 & 20 & 30 & 64\\
	cumulative time [h] & 10 & 30 & 60 & 124\\
\end{tabular}
\end{ruledtabular}
\end{table}

\subsection{Structure and chemical composition}
In order to determine the microscopic structure of the coating samples, as well as their change upon annealing, we performed a series of grazing-incidence X-ray diffraction (GI-XRD) measurements with a Philips MRD diffractometer, equipped with a Cu tube operated at 40 kV and 40 mA. The probe beam was collimated and partially monochromatized to the Cu K-$\alpha$ line by a parabolic multilayer mirror, whereas the detector was equipped with a parallel plate collimator to define the angular acceptance.

Rutherford back-scattering spectrometry (RBS) and elastic recoil detection with time-of-flight detection (ERD-TOF) \cite{Chicoine17} were used to determine the composition of the coating samples, after deposition and after the different annealing steps. RBS measurements were carried out using $^4$He beams: at 2 MeV in order to rely on the Rutherford cross section of O and F, and at 3.7 MeV to better resolve the different elements. The beam was incident at an angle of 7$^\circ$ from the normal, and the detector was placed at a scattering angle of 170$^\circ$. For ERD-TOF, a 50 MeV Cu beam was incident at 15$^\circ$ from the sample surface and the TOF camera was at 30$^\circ$ from the beam axis.

\subsection{Optical properties}
\label{SECT_opt}
We used two J. A. Woollam spectroscopic ellipsometers to measure the coating optical properties and thickness, covering complementary spectral regions from ultraviolet to infrared: a VASE for the 0.73-6.53 eV photon energy range (corresponding to a 190-1700 nm wavelength range) and a M-2000 for the 0.74-5.06 eV range (245-1680 nm). The coated Si wafers were measured in reflection, their complex reflectance ratio was characterized by measuring its amplitude component $\Psi$ and phase difference $\Delta$ \cite{Fujiwara07}. To maximize the response of the instruments, ($\Psi$, $\Delta$) spectra were acquired at different incidence angles ($\theta$ = 50$^\circ$, 55$^\circ$, 60$^\circ$) close to the coating Brewster angle. Coating refractive index and thickness were derived by fitting the spectra with realistic optical models \cite{Fujiwara07}. The optical response of the bare  Si wafers had been characterized with prior dedicated measurements. Further details about our ellipsometric analysis are available elsewhere \cite{Amato19}.

We used photo-thermal deflection \cite{Boccara80} to measure the coating optical absorption at $\lambda=1064$ nm with an accuracy of less than 1 part per million (ppm).

\subsection{Mechanical properties}
\label{SECTmethMech}
Two nominally identical fused-silica disks, named A and B, were used for the characterization of the coating mechanical properties. We measured their mass with an analytical balance, before and after each treatment (coating deposition, annealing runs), and their diameter with a caliper. We then used the measured coated area, coating thickness from ellipsometric measurements and mass values to calculate the coating density $\rho$.

We used the ring-down method \cite{Nowick72} to measure the frequency $f$ and ring-down time $\tau$ of the first vibrational modes of each fused-silica disk, before and after the coating deposition, and calculated the coating loss angle
\begin{equation}
\label{EQcoatLoss}
\varphi_c = \frac{\varphi + (D-1)\varphi_0}{D} \ ,
\end{equation}
where $\varphi_0 = (\pi f_0 \tau_0)^{-1}$ is the measured loss angle of the bare substrate, $\varphi = (\pi f \tau)^{-1}$ is the measured loss angle of the coated disk. $D$ is the frequency-dependent measured \textit{dilution factor} \cite{Li14},
\begin{equation}
\label{EQdilFact}
D = 1 -  \frac{m_0}{m} \left( \frac{f_0}{f} \right)^2 \ ,
\end{equation}
where $m_0$, $m$ is the disk mass as measured before and after the coating deposition, respectively.

\begin{figure}
\centering
	\includegraphics[width=0.30\textwidth]{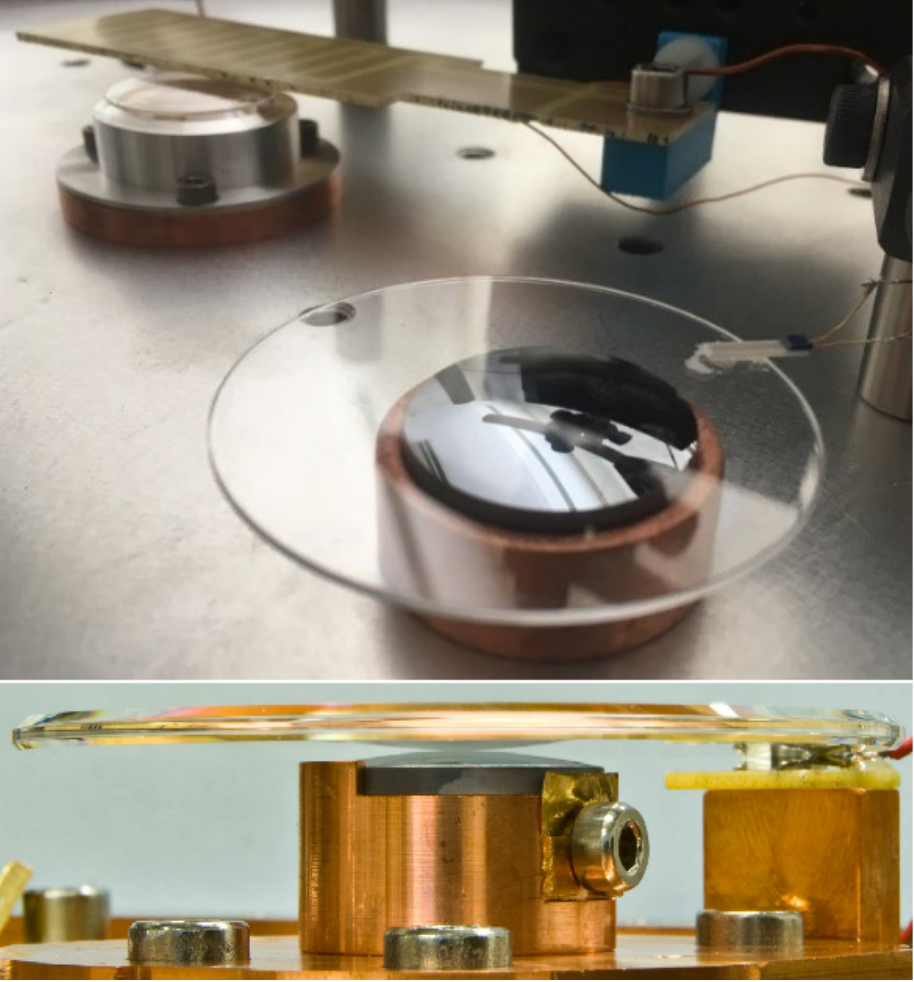}
	\caption{GeNS systems used at Universit\`{a} degli Studi di Urbino Carlo Bo (top) and at Laboratoire des Mat\'{e}riaux Avanc\'{e}s (bottom) to measure the mechanical properties of thin films.}
	\label{FIG.GeNSs}
\end{figure}

We measured modes from $\sim$2.5 to $\sim$39 kHz for each fused-silica disk, in a frequency band which partially overlaps with the detection band of ground-based gravitational-wave detectors ($10 - 10^4$ Hz). In order to avoid systematic damping from suspension and ambient pressure, we used two clamp-free in-vacuum Gentle Nodal Suspension (GeNS) systems \cite{Cesarini09}, shown in Fig. \ref{FIG.GeNSs}. This kind of system is currently the preferred solution of the Virgo and LIGO Collaborations for performing internal friction measurements \cite{Granata20,Vajente17}.

The fused-silica disks were first measured at LMA before and after coating deposition, then measured, annealed and measured again at Universit\`{a} degli Studi di Urbino Carlo Bo (UniUrb). After deposition, coating Young modulus $Y$ and Poisson ratio $\nu$ were estimated by fitting finite-element simulations to the measured dilution factor via least-squares numerical regression where we used values of substrate thickness previously determined by fitting the measured mode frequencies with a specific sub-set of simulations \cite{Granata20}. Further details about our GeNS systems, finite-element simulations and data analysis are available elsewhere \cite{Granata20,Granata16}.

\section{Results}

\subsection{Structure and chemical composition}
\label{SECT_struct_comp}
\begin{figure}
\centering
	\includegraphics{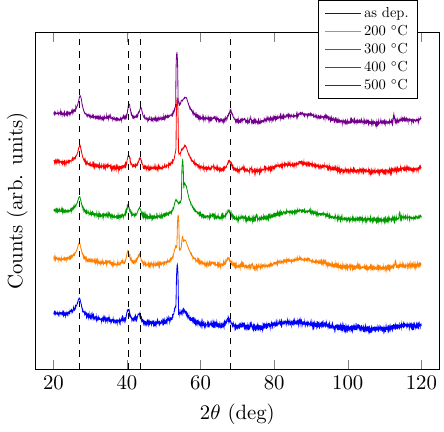}
	\caption{GI-XRD diffractograms of IBS MgF$_2$ thin films on silicon wafers, acquired before and after annealing at different soaking temperatures $T_a$. Diffraction peaks at about 27$^\circ$, 40$^\circ$, 44$^\circ$ and 68$^\circ$ fairly match the $2\theta$ values expected for a crystalline tetragonal structure (JCPDS 70-2269, vertical dashed lines) \cite{Quesnel00,Ristau02,Yu07}, peaks between 50$^\circ$ and 60$^\circ$ are mainly due to the background signal of the silicon substrate.}
	\label{FIG.XRD}
\end{figure}

The GI-XRD diffractograms of the coating samples are shown on Fig. \ref{FIG.XRD}, where it can be seen that diffraction peaks at about 27$^\circ$, 40$^\circ$, 44$^\circ$ and 68$^\circ$ are already present in the as-deposited coating. Those peaks fairly match the $2\theta$ values expected for a crystalline tetragonal structure (JCPDS 70-2269) \cite{Quesnel00,Ristau02,Yu07}. Other peaks between 50$^\circ$ and 60$^\circ$ are mainly due to the background signal of the silicon substrate. The change of coating peaks was minimal for the annealed samples, as they became just slightly higher and narrower. The presence of a poly-crystalline phase in the coatings is particularly relevant for gravitational-wave detectors, since it usually is a source of scattered light and hence of optical loss and noise.

Results of the RBS measurements are listed in Table \ref{TABLE_RBS}. Relative atomic concentrations and density $\rho$ were deduced from SIMNRA simulations \cite{Mayer99} of the RBS spectrum acquired on each sample. The Mg/F concentration ratio is compatible with 0.5 for all samples, before and after annealing, within the measurement uncertainty. In addition, all samples contain 3-5\% O and 0.4-0.5\% H, and are contaminated by the sputtering gas (0.5-0.8\% Xe) and by Mo from the sputtering source grids. The Mo content increases from about 0.4\% at the substrate interface to 0.7\% near the surface. All the samples also contain traces of Cu, Ar and Ta ($<0.1$\%). The areal atomic density found with RBS can be divided by the layer thickness measured via spectroscopic ellipsometry (206 nm), to find a coating density $\rho$ close to 3.0 g/cm$^3$ for all samples. According to our analisys, the sample annealed at $T_a=500$ $^\circ$C also featured a 3.7 nm thick top layer of MgO, assuming an MgO bulk density of 3.85 g/cm$^3$; hence, this sample apparently suffered from some surface degradation due to oxidation.
\begingroup
\squeezetable
\begin{table*}
\caption{\label{TABLE_RBS} Relative atomic concentrations (\%) and density $\rho$ of IBS MgF$_2$ thin films, before and after annealing at different soaking temperatures $T_a$, deduced from SIMNRA simulations \cite{Mayer99} of the RBS spectrum acquired on each sample. The Mg/F ratio is in at./at. and the density was calculated by assuming a layer thickness of 206 nm for all samples, as measured via spectroscopic ellipsometry on as-deposited samples. ($^{*}$)Mo concentration decreases with depth in all samples, from about 0.7\% at the surface to about 0.45\% at the substrate interface; average values are shown. ($^{**}$)The sample annealed at $T_a=500$ $^\circ$C features an MgO surface layer $\sim$4 nm thick.}
\begin{ruledtabular}
\begin{tabular}{lccccccccccccc}
	& Mg & F & O & H & Ar & Cu & Mo($^{*}$) & Xe & Ta & Al & Mg/F & $\rho$ [g/cm$^3$]\\ \cline{2-13}
	as deposited & 32.2 & 63 & 3 & 0.4 & 0.04 & 0.05 & 0.53 & 0.52 & 0.01 & 0.6 & 0.51 & 2.96 $\pm$ 0.05\\
	200 $^\circ$C & 31.5 & 62 & 5 & 0.5 & 0.04 & 0.05 & 0.53 & 0.71 & 0.01 & & 0.51 & 2.97 $\pm$ 0.05\\
	300 $^\circ$C & 31.1 & 62 & 5 & 0.5 & 0.04 & 0.05 & 0.60 & 0.69 & 0.01 & 0.5 & 0.50 & 3.00 $\pm$ 0.05\\
	400 $^\circ$C & 30.8 & 62 & 5 & 0.5 & 0.08 & 0.05 & 0.56 & 0.60 & 0.01 & & 0.49 & 2.98 $\pm$ 0.05\\
	500 $^\circ$C ($^{**}$) & 30.4 & 63 & 5 & 0.4 & 0.04 & 0.05 & 0.61 & 0.84 & 0.01 & & 0.48 & 2.83 $\pm$ 0.05\\
\end{tabular}
\end{ruledtabular}
\end{table*}
\endgroup

\subsection{Optical properties}
\label{SECT_opt_prop}
\begin{figure*}
\centering
	\includegraphics{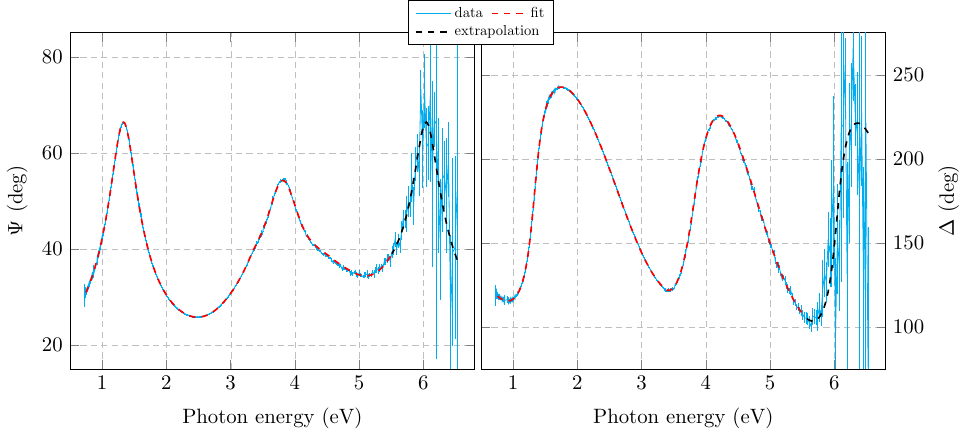}
	\caption{Measured ellipsometric spectra of IBS MgF$_2$ thin films, acquired at an incidence angle $\theta$ = 60$^\circ$.}
	\label{FIG.spectra}
\end{figure*}
By way of example, figure \ref{FIG.spectra} shows $(\Psi,\Delta)$ spectra of the as-deposited coatings, acquired at an incidence angle $\theta$ = 60$^\circ$. As the band gap of crystalline magnesium fluoride is 10.8 eV \cite{Babu11}, we initially expected MgF$_2$ coatings to be transparent in the energy region probed by our ellipsometers. Instead, preliminary measurements showed that some optical absorption in the ultraviolet region had to be taken into account, in order to explain the observed degradation of data quality above 5.7 eV, where the signal to noise ratio was drastically reduced, and to correctly fit our data. Such absorption could be explained by the presence of color centers \cite{Quesnel00,Dumas02}, as well as by the observed 0.5-0.7\% Mo contamination or the O-related centers due to the 5\% O in the samples. We then used a two-pole function and a Tauc-Lorentz oscillator for the optical model of the thin films, which better reproduced the data and simultaneously fitted all the measured spectra with the same accuracy. In particular, the pole in the ultraviolet region takes into account absorption at higher photon energy which affects the real part of the dielectric function in the measurement region, and the pole in the infrared region allows the refractive index to have an inflection point. The Tauc-Lorentz model describes the optical absorption, but the exact energy of the oscillator could not be accurately determined, due to the poor data quality in the ultraviolet region. However, for the same reason, data for $E > 5.7$ eV had a negligible influence on the fit algorithm, and the results were compatible with those obtained by fitting the data up to 5.5 eV and extrapolating the values to higher photon energies.

\begin{figure}
\centering
	\includegraphics{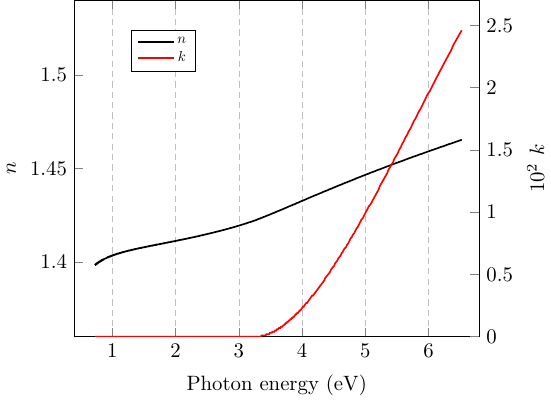}
	\caption{Refractive index $n$ and extinction coefficient $k$ of as-deposited IBS MgF$_2$ thin films as a function of photon energy, derived from ellipsometric measurements. Relevant values for present and future gravitational-wave detectors are 0.80 and 1.17 eV, corresponding to a laser wavelength of 1550 and 1064 nm, respectively. For energy values smaller than $\sim$3.5 eV, the extinction is smaller than the sensitivity of the ellipsometers ($k < 10^{-3}$).}
	\label{FIG.optConst}
\end{figure}

Figure \ref{FIG.optConst} shows the dispersion law and the extinction curve derived from our analysis of the as-deposited coating data, and Table \ref{TABLEoptProp} lists our results against those we found in the literature concerning IBS MgF$_2$ thin films \cite{Allen90,Bosch00,Quesnel00,Gunster05,Yoshida06}. Values at $E$ = 1.17 eV and $E$ = 0.80 eV are particularly relevant, since those photon energies correspond to 1064 and 1550 nm, respectively, which are the operational laser wavelenghts of current and future gravitational-wave detectors \cite{aLIGO,AdVirgo,KAGRA,ET1,ET2}. Refractive index values are $n = 1.405 \pm 0.005$ at 1064 nm and $n = 1.401 \pm 0.005$ at 1550 nm. For comparison, the refractive index at 1064 nm of the IBS silica coatings of present detectors is $n = 1.47 \pm 0.01$ before annealing \cite{Granata20}. Extinction at 6.4 eV (193 nm) is considerably higher than the one reported in the literature \cite{Gunster05,Yoshida06}, due to the high absorption we observed in the ultraviolet region of the spectra.
\begin{figure}
\centering
	\includegraphics{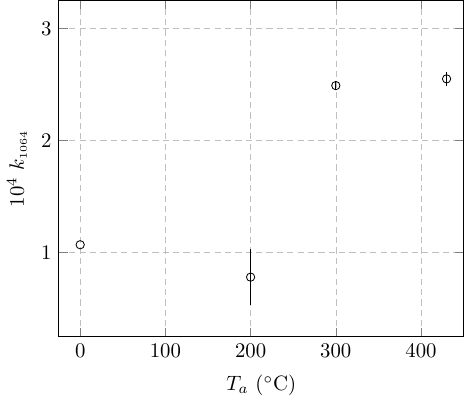}
	\caption{Extinction coefficient $k_{\textrm{\tiny{1064}}}$ of IBS MgF$_2$ thin films as a function of the annealing temperature $T_a$, obtained from photo-thermal deflection measurements of optical absorption performed at 1064 nm ($T_a = 0$ $^{\circ}$C denotes as-deposited coatings).}
	\label{FIG.absorptVSann}
\end{figure}
\begingroup
\squeezetable
\begin{table*}
\caption{\label{TABLEoptProp} Refractive index $n$, extinction coefficient $k$ and density $\rho$ of as-deposited IBS MgF$_2$ thin films. $n$ and $k$ values presented in this work were measured in the 190-1680 nm wavelength range via spectroscopic ellipsometry. The $k$ value at 1064 nm was deduced from photo-thermal deflection measurements of optical absorption, by assuming negligible scatter loss. $^{(*)}$Extrapolations.}
\begin{ruledtabular}
\begin{tabular}{ccccccccc}
	$E$ [eV] & $\lambda$ [nm] & & This work & Allen et al. \cite{Allen90} & Bosch et al. \cite{Bosch00} & Quesnel et al. \cite{Quesnel00} & G\"{u}nster et al. \cite{Gunster05} & Yoshida et al. \cite{Yoshida06}\\ \hline
	0.80 & 1550 & $n$ & 1.401 $\pm$ 0.005 & & 1.380 $^{(*)}$\\
	& & $k$ & $<10^{-3}$ & & $1.5\times 10^{-5}$ $^{(*)}$\\
	0.94 & 1320 & $n$ & 1.403 $\pm$ 0.005 & & 1.380 $^{(*)}$\\
	& & $k$ & $<10^{-3}$ & $7 \times 10^{-4}$ & $1.7\times 10^{-5}$ $^{(*)}$\\
	1.17 & 1064 & $n$ & 1.405 $\pm$ 0.005 & & 1.380 $^{(*)}$\\
	& & $k$ & $(1.062 \pm 0.004)\times 10^{-4}$ & $5 \times 10^{-4}$ & $2.0\times 10^{-5}$ $^{(*)}$\\
	1.96 & 633 & $n$ & 1.411 $\pm$ 0.005 & 1.453 $\pm$ 0.023 & 1.383\\
	& & $k$ & $<10^{-3}$ & & $4.2\times 10^{-5}$\\
	3.53 & 351 & $n$ & 1.426 $\pm$ 0.005 & & 1.391 & $1.390 - 1.41$ \\
	& & $k$ & $<10^{-3}$ & & $1.7\times 10^{-4}$ & $8 \times 10^{-6} - 3.3 \times 10^{-2}$\\
	6.42 & 193 & $n$ & 1.46 $^{(*)}$ & & & & $1.44 - 1.45$ & 1.44\\
	& & $k$ & 0.024 $^{(*)}$ & & & & $2-5\times 10^{-3}$ & $2\times 10^{-4}$\\
	& & $\rho$ [g/cm$^{3}$] & 2.7 $\pm$ 0.2 & 3.18\\
\end{tabular}
\end{ruledtabular}
\end{table*}
\endgroup

Figure \ref{FIG.absorptVSann} shows the extinction coeffient $k$ obtained from the photo-thermal deflection measurements of optical absorption at 1064 nm, as a function of the annealing temperature $T_a$. We assumed that loss by light scatter was negligible. We obtained $k = 1.1 \times 10^{-4}$ before treatment, which is about three orders of magnitude larger than that of the as-deposited silica layers of current gravitational-wave detectors. Although we expect the extinction to be approximately of the same order of magnitude at longer wavelengths, work is currently ongoing to upgrade our apparatus in order to perform sensitive measurements also at 1550 and possibly 2000 nm, which are relevant wavelengths for future detectors \cite{ET1,ET2,Abbott17}.

The first annealing step at $T_a=200$ $^\circ$C decreased the extinction by 27\%, but subsequent treatments at higher temperature considerably increased it. Thus, the annealing temperature for minimum extinction due to optical absorption is between $200$ and 300 $^\circ$C.

\subsection{Mechanical properties}
The main features of fused-silica disks A and B used for the measurements are presented in Table \ref{TABLEsamples}.

Beside the fact of providing a cross-check of the results, the use of two independent GeNS systems allowed us to identify and correct for a systematic effect due to the sample temperature, as described in the following.
\begin{table}
\caption{\label{TABLEsamples} Coated fused-silica disks used to characterize the coating mechanical properties: diameter $\varnothing$, substrate thickness $d_0$, mass $m_0$ before coating, mass $m$ after coating, coating thickness $d$ on each side.}
\begin{ruledtabular}
\begin{tabular}{lcc}
	& A & B\\
	$\varnothing$ [mm] &  49.77 $\pm$ 0.03 & 49.92 $\pm$ 0.01\\
	$d_0$ [mm]	&	1.09 $\pm$ 0.01 &	1.08 $\pm$ 0.01\\
	$m_0$ [g] &  4.6158 $\pm$ 0.0001 & 4.6348 $\pm$ 0.0001\\
	$m$ [g] &  4.6180 $\pm$ 0.0001 & 4.6369 $\pm$ 0.0004\\
	$d$ [nm] &  206 $\pm$ 2 & 206 $\pm$ 2\\
\end{tabular}
\end{ruledtabular}
\end{table}

\begin{figure}
\centering
	\includegraphics[width=0.50\textwidth]{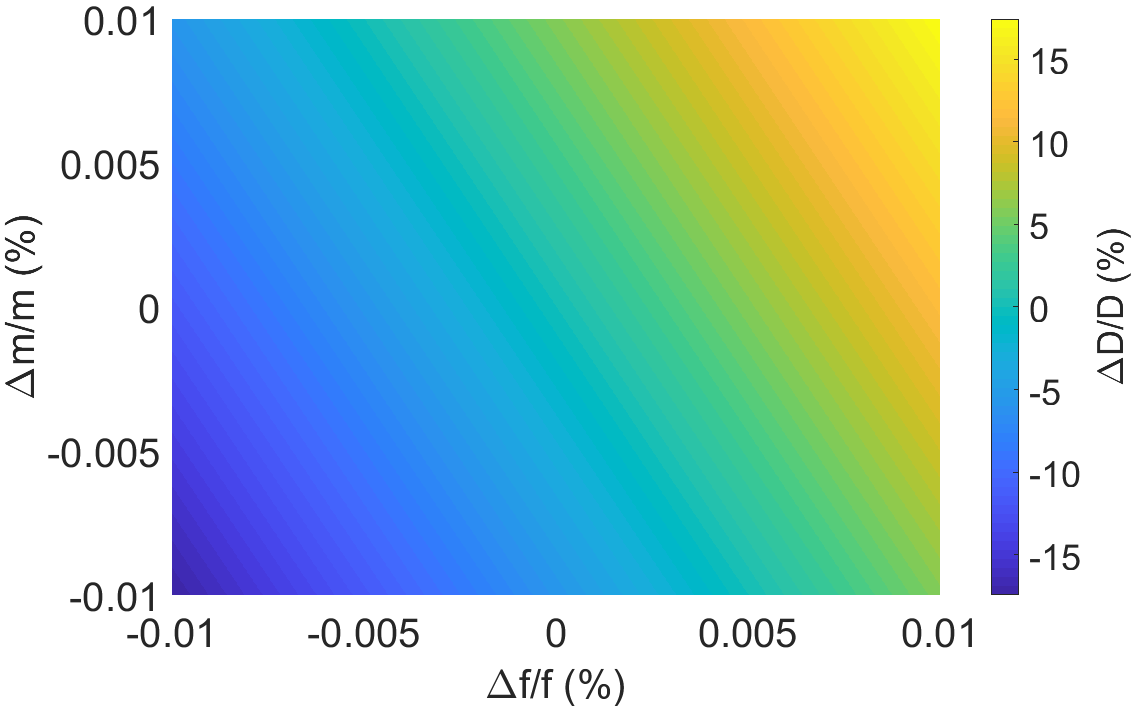}
	\caption{Relative error $\Delta D/D$ on dilution factor as a function of relative errors $\Delta f/f$ and $\Delta m/m$ on frequency and mass, respectively, for disk B ($f_0 = 2681.062$ Hz, $f = 2682.759$ Hz). See Eq. (\ref{EQdilFact}) and Table \ref{TABLEsamples} for more details.}
	\label{FIG.deltaD}
\end{figure}

\begin{figure}
\centering
	\includegraphics{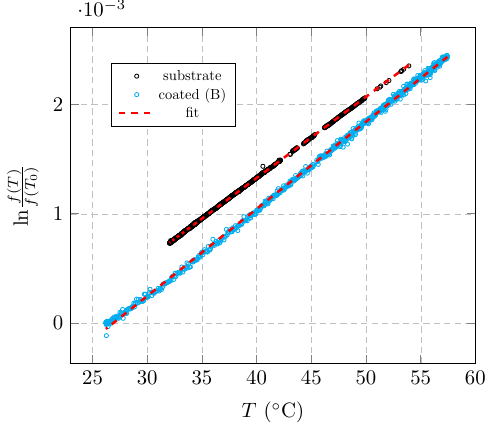}
	\caption{Variation of mode frequency $f(T)$ as a function of sample temperature $T$, for a fused-silica bare disk and coated disk B.}
	\label{FIG.freqFit}
\end{figure}

By definition, the measured dilution factor $D$ is very sensitive to variations of frequencies and masses. As shown by Fig. \ref{FIG.deltaD}, $\Delta D/D$ can be as high as $\sim$15\% if $\Delta f/f$ and $\Delta m/m$ are both of the order of 0.01\%. Indeed, the frequency ratio in Eq.(\ref{EQdilFact}) depends on the Young modulus of the sample, which is in turn temperature dependent. For temperatures close to or higher than 300 K and in a limited temperature range, the relative variation of the sample Young modulus with temperature is a constant \cite{Wachtman61}, $\eta = (dY/dT)/Y$, whereas the sample mode frequencies are proportional to the square root of the Young modulus, $f \propto \sqrt{Y}$ \cite{Amabili95}. Thus we expect that
\begin{equation}
\ln \frac{f(T)}{f(T_0)} =\frac{\eta}{2}\left (T-T_0\right ) \ ,
\end{equation}
where $f(T)$ is the mode frequency at temperature $T$. 
Our GeNS system at LMA is installed in a clean room where the temperature is stabilized to (21.9 $\pm$ 0.5) $^\circ$C, while our GeNS system at UniUrb is in a room without temperature control. Each setup has a temperature probe in its vacuum tank: right under the GeNS copper base plate at LMA, on a twin suspended sample\footnote{We found no experimental evidence that the laser of the optical lever used to measure the ring-down amplitude of the main sample induced a temperature variation, therefore we assumed that the temperature measured on the twin sample is equal to that of the main sample.} at UniUrb (visible in the foreground of Fig. \ref{FIG.GeNSs}). In order to measure the change of resonant frequencies with temperature, we installed heating strips around the vacuum tank of the GeNS system at UniUrb and slowly heated a fused-silica bare disk, monitoring the frequency of its first mode. That bare disk was nominally identical to disks A and B, from their same batch. Afterward, we applied the same procedure also to the coated disk B. Figure \ref{FIG.freqFit} shows the results of those measurements. We obtained $\eta = (1.50 \pm 0.01) \times 10^{-4}$ $^{\circ}$C$^{-1}$ for the bare disk and $\eta = (1.58 \pm 0.01) \times 10^{-4}$ $^{\circ}$C$^{-1}$ for coated disk B, by linearly fitting the data in a semi-logaritmic scale. We then used these values to perform a correction of measured mode frequencies by an amount
\begin{equation}
\label{EQ_freqCorr}
\Delta f = \frac{\eta}{2} (T-T_0)\ f(T_0) 
\end{equation}
for data of both fused-silica disks A and B. This correction is critical, whenever mode frequencies are measured in a system where temperature may drift.

\begin{figure*}
\centering
	\includegraphics{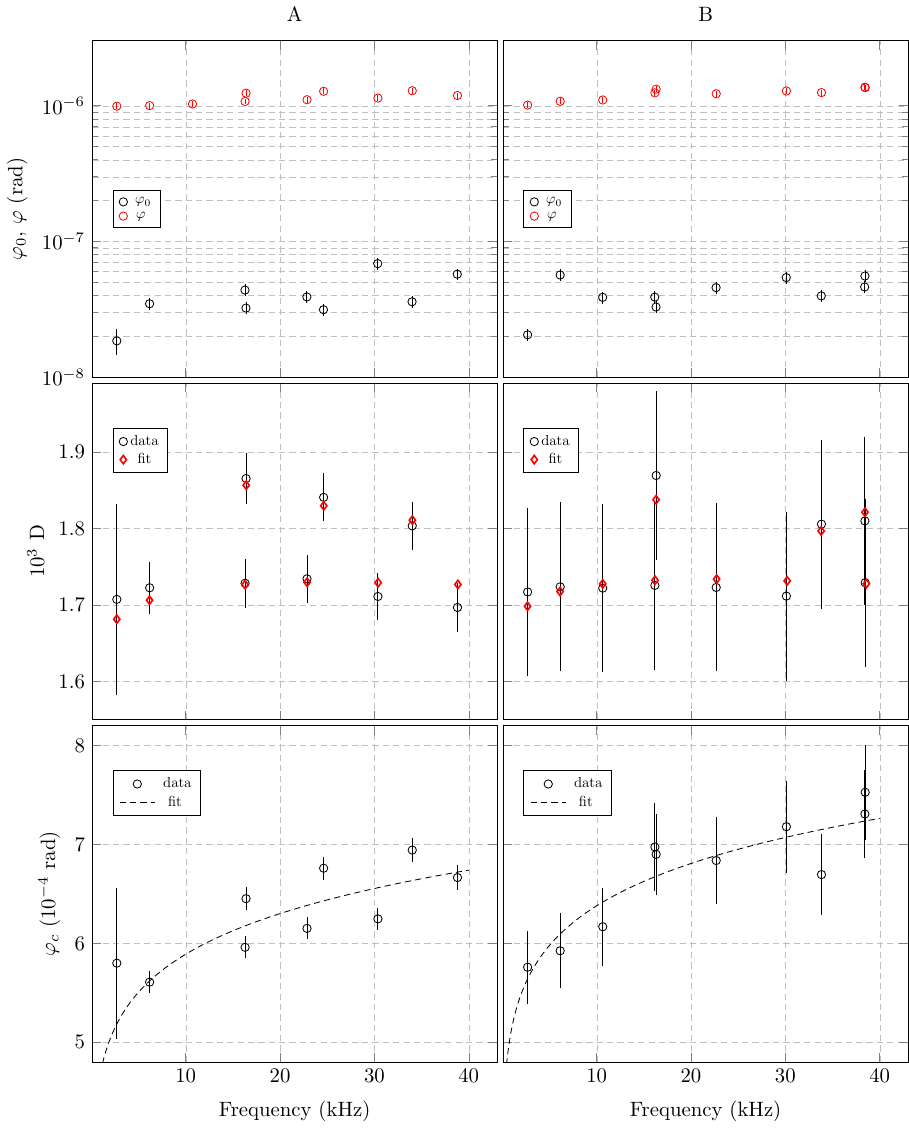}
	\caption{Characterization of loss angles of fused-silica disks A ({\it left column}) and B ({\it right column}), as a function of frequency. {\it Top row:} measured loss angles before and after deposition of IBS MgF$_2$ thin films ($\varphi_0$ and $\varphi$, respectively). {\it Middle row:} comparison between measured and least-squares best-fit simulated dilution factor $D$. {\it Bottom row:} coating loss angle $\varphi_c$ of as-deposited IBS MgF$_2$ thin films; the best-fit power-law model of Eq.(\ref{EQpowerLaw}) is also shown (dashed line). See Eq.(\ref{EQcoatLoss}) for more details.}
	\label{FIG.lossLMA}
\end{figure*}

Figure \ref{FIG.lossLMA} shows the dilution factor and loss angles of fused-silica disks A and B, as measured at LMA. In the 2.5--39 kHz frequency band, the mechanical loss of the as-deposited IBS MgF$_2$ coatings ranges from about $5.5 \times 10^{-4}$ to $7.5 \times 10^{-4}$ rad, that is, 20 to 30 times higher than that of the as-deposited silica layers of current gravitational-wave detectors \cite{Granata20}. This excess loss might be partly explained by the poly-crystalline phase of the MgF$_2$ coatings. 

For comparison, Kinbara et al. measured a coating loss angle of $3.5 \times 10^{-4}$ to $5 \times 10^{-4}$ rad at 30 Hz on thermally evaporated MgF$_2$ thin films \cite{Kinbara81,Kinbara82}. Such different values could be explained by the different frequency of their measurement, and possibly also by the different nature of their thin films, grown with a different technique.

In order to describe the observed frequency-dependent behavior of the coating loss angle, we fit a power-law model \cite{Gilroy81,Travasso07,Cagnoli18}
\begin{equation}
\label{EQpowerLaw}
\varphi_c(f) = a\left(\frac{f}{10 \textrm{ kHz}}\right)^b
\end{equation}
to our data via least-squares linear regression. Table \ref{TABLEmechProp} lists the best-fit parameters $(a,b)$ for each measured fused-silica disk, together with the best-fit estimations of coating Young modulus $Y$ and Poisson ratio $\nu$ obtained via the dilution factor fitting procedure described in Section \ref{SECTmethMech}. By taking the average of the results obtained with the two coated disks A and B, we obtain $Y = 115$ GPa and $\nu = 0.27$. Kinbara et al. \cite{Kinbara81,Kinbara82} obtained $Y = 70$ GPa and $Y = 150$ GPa by applying the resonant method to two different substrates, but could not identify the reason for such discrepancy. Our results fall within that range.

For the as-deposited samples, we obtained a density $\rho$ of 2.7 $\pm$ 0.2 g/cm$^{3}$ from coating mass and thickness ellipsometric measurements, a value fairly close but lower than that of 2.96 g/cm$^{3}$ obtained via RBS. Similarly to the coating sample annealed at 500 $^\circ$C, which has the lowest density as measured through RBS ($\rho = 2.83$ g/cm$^{3}$), this might be explained by the fact that the samples used for the characterization of the coating mechanical properties suffered from some surface oxidation, despite their storage under primary vaccum.
\begin{table}
\caption{\label{TABLEmechProp} Measured mechanical properties of as-deposited IBS MgF$_2$ thin films: Young modulus $Y$, Poisson ratio $\nu$ and best-fit parameters of the power-law model of Eq.(\ref{EQpowerLaw}) used to describe the observed frequency-dependent behavior of the coating loss angle.}
\begin{ruledtabular}
\begin{tabular}{ccccc}
	& $Y$ [GPa] & $\nu$ & $a$ [$10^{-4}$ rad] & $b$\\ 
	disk A & 115 $\pm$ 3 & 0.28 $\pm$ 0.02 & 6.4 $\pm$ 0.1 & 0.09 $\pm$ 0.02\\
	disk B & 115 $\pm$ 3 & 0.26 $\pm$ 0.02 & 5.9 $\pm$ 0.2 & 0.10 $\pm$ 0.03\\
\end{tabular}
\end{ruledtabular}
\end{table}

\begin{figure*}
\centering
	\includegraphics{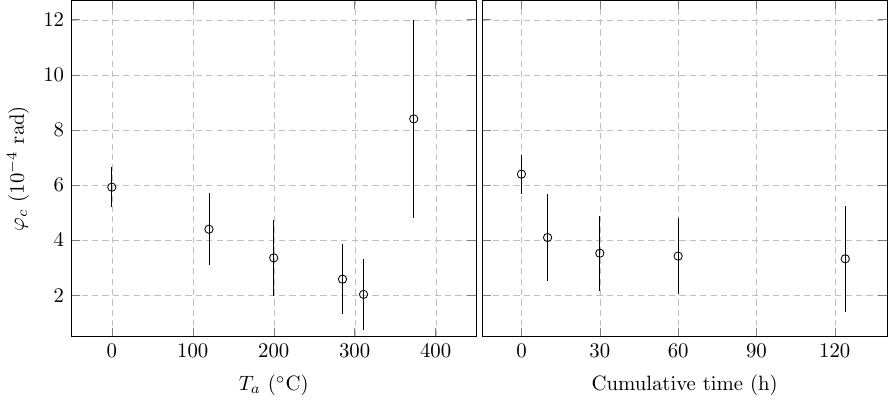}
	\caption{Coating loss angle $\varphi_c$ of IBS MgF$_2$ thin films, as a function of annealing temperature $T_a$ for a soaking time $\Delta t_a = 10$ h ({\it left}) and of cumulative soaking time at temperature $T_a = 285$ $^{\circ}$C ({\it right}), see Tables \ref{TABLEannA} and \ref{TABLEannB} for more details ($T_a = 0$ $^{\circ}$C denotes as-deposited coatings).}
	\label{FIG_lossUniUrb}
\end{figure*}

\begin{figure}
\centering
	\includegraphics[width=0.45\textwidth]{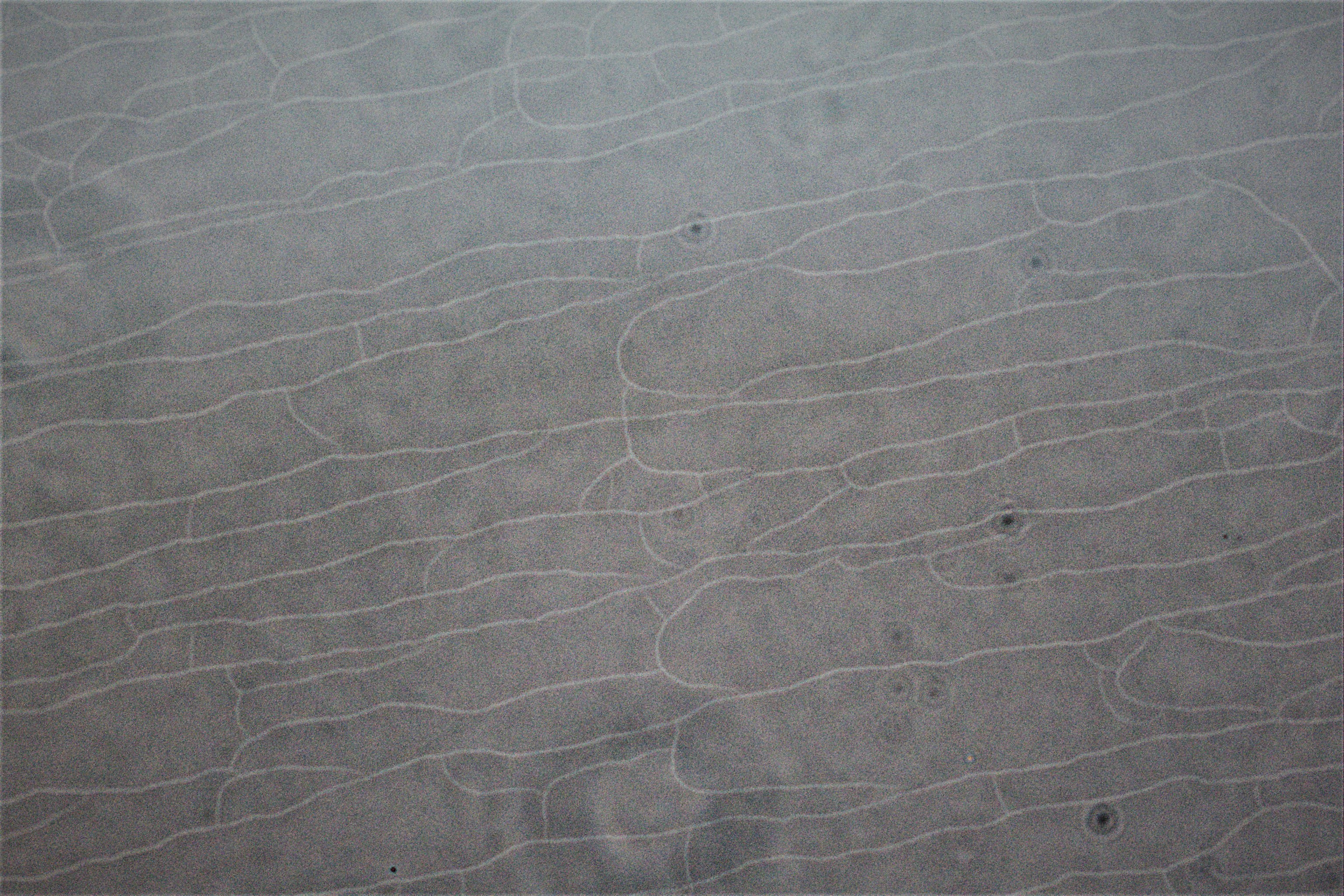}
	\caption{Cracks at the surface of the IBS MgF$_2$ thin films on disk A, observed after annealing at $T_a = 373$ $^\circ$C.}
	\label{FIG.cracks}
\end{figure}

Figure \ref{FIG_lossUniUrb} shows the effect of the post-deposition annealing treatments on the average coating loss angle of disks A and B, calculated from several exemplary modes at different frequencies. For this data, acquired without temperature stabilization at UniUrb, we applied the frequency correction described by Eq.(\ref{EQ_freqCorr}). As we increased the annealing temperature up to $T_a = 311$ $^\circ$C, the average coating loss of disk A monotonically decreased from the initial value of $(5.9\pm0.7) \times 10^{-4}$ rad to $(2.0 \pm 1.3) \times 10^{-4}$ rad. After treatment at $T_a = 373$ $^\circ$C, however, its average coating loss increased to $(8.4 \pm 3.6) \times 10^{-4}$ rad. Such substantial increase might be explained by the appearence of cracks on the coating surface, observed on disk A with an optical microscope and shown on Fig. \ref{FIG.cracks}, likely due to the fact that the SiO$_2$ substrate and the MgF$_2$ coatings have different thermal expansion coefficients. Similar cracks were previously observed by Kinbara et al., and ascribed to the relaxation of accumulated stress \cite{Kinbara81,Kinbara82}. Regardless, the annealing temperature for minimum coating loss angle $\varphi_c$ is around $T_a = 311$ $^\circ$C.

Concerning the annealing time, in order to avoid the formation of cracks, we used a soaking temperature $T_a = 285$ $^\circ$C for our tests. The average coating loss angle of disk B decreased after each step until the cumulative time of treatment amounted to 30 hours, when it was $(3.5 \pm 1.4) \times 10^{-4}$ rad. However, the change in coating loss between the step of 10 hours and those of longer cumulative soaking time is negligible, when compared to the measurement uncertainty. Thus, in summary, we found that a soaking time longer than 10 hours has no effect on the average coating loss angle value, for $T_a = 285$ $^\circ$C.

\section{Conclusions}
In the framework of a research activity devoted to find low-noise coating materials for present and future gravitational-wave detectors \cite{Granata20review}, we characterized the optical and mechanical properties of a set of IBS MgF$_2$ thin films. We chose fluoride coatings because of their low refractive index $n_{\textrm{\tiny{L}}}$, with the aim of minimizing the overall high-reflection coating thickness $d$ in Eq.(\ref{eqn.S}). As a reminder, $d$ is a monotonically decreasing function of the refractive index contrast $C = n_{\textrm{\tiny{H}}}/n_{\textrm{\tiny{L}}}$. Furthermore, because of their potentially low mechanical loss at low temperature \cite{Schwarz11}, fluorides could be a valid option for use in cryogenic detectors.

Indeed, the IBS MgF$_2$ thin films featured a 4\% lower refractive index than that of IBS silica layers of current detectors \cite{Granata20}, at 1064 nm. However, their optical absorption and ambient-temperature loss angle turned out to be considerably higher, likely because they were partially poly-crystalline. In order to minimize such losses, the coating samples were thermally treated with increasing soaking temperature and time. A soaking temperature of $T_a = 285$ $^{\circ}$C avoided the formation of cracks and minimized the coating loss angle value, while the lowest optical absorption occurred after thermal treatment at $T_a = 200$ $^{\circ}$C. As a consequence, the optimal soaking temperature for our set of samples proved to be between 200 and 300 $^{\circ}$C, where both the coating optical absorption and average loss angle were close to their minimum values. Soaking times longer than 10 hours had a negligible effect on the average coating loss angle value.

However, regardless of the effects of annealing, the implementation of IBS MgF$_2$ thin films in gravitational-wave detectors would require their optical absorption to be reduced drastically, by at least by 3 orders of magnitude. Similarly, their ambient-temperature loss angle also proved to be too large by at least one order of magnitude. Lower optical absorption and loss angle could possibly be achieved by changing the coating growth conditions \cite{Granata20}, as well as by reducing the amount of impurities. As shown in Table \ref{TABLEoptProp}, for instance, Bosch et al. and Quesnel et al. demonstrated that IBS MgF$_2$ thin films of significantly lower extinction and refractive index can be produced \cite{Bosch00,Quesnel00}. The poly-crystalline phase of the as-deposited coatings, which is usually source of scattered light, might in principle be avoided by using different growth conditions as well.

The optimization of growth parameters, together with the measurement of the low-temperature mechanical loss angle and of optical absorption at longer wavelengths, will be the object of future studies.

\section*{Acknowledgments}
This work has been promoted by the Laboratoire des Mat\'{e}riaux Avanc\'{e}s and partially supported by the Virgo Coating Research and Development (VCR\&D) Collaboration. The work carried out at U. Montr\'{e}al is supported by the Fonds de Recherche du Qu\'{e}bec - Nature et Technologie (FRQNT) through the Regroupement Qu\'{e}becois sur les Matériaux de Pointe (RQMP) on equipment obtained in part thanks to the Canada Foundation for Innovation (CFI) and the Natural Sciences and Engineering Research Council (NSERC). The authors would like to thank M. Gauch, F. Carstens and H. Ehlers of the Laser Zentrum Hannover for the production of the MgF$_2$ thin films and for the fruitful discussions, as well as M. Fazio for the first and accurate review of the manuscript. In the online document repositories of the
LIGO and the Virgo Scientific Collaborations, this work has been assigned document numbers LIGO-P2100113 and VIR-0314D-21.

\end{document}